\documentclass[twocolumn,showpacs,preprintnumbers,amsmath,amssymb]{revtex4}

\usepackage{graphicx}
\usepackage{dcolumn}
\usepackage{bm}

\begin{document}

\title{Heating-compensated constant-temperature tunneling measurements on stacks of
Bi$_2$Sr$_2$CaCu$_2$O$_{8+x}$ intrinsic junctions}


\author{Myung-Ho Bae}
\author{Jae-Hyun Choi}
\author{ Hu-Jong Lee}%

\affiliation{Department of Physics, Pohang University of Science
and Technology, Pohang 790-784, Republic of Korea}%

\date{\today}

\begin{abstract}
In highly anisotropic layered cuprates such as
Bi$_2$Sr$_2$CaCu$_2$O$_{8+x}$ tunneling measurements on a stack of
intrinsic junctions in a high-bias range are often susceptible to
self-heating. In this study we monitored the temperature variation
of a stack (``sample stack") of intrinsic junctions by measuring
the resistance change of a nearby stack (``thermometer stack") of
intrinsic junctions, which was strongly thermal-coupled to the
sample stack through a common Au electrode. We then adopted a
proportional-integral-derivative scheme incorporated with a
substrate-holder heater to compensate the temperature variation.
This in-situ temperature monitoring and controlling technique
allows one to get rid of spurious tunneling effects arising from
the self-heating in a high bias range.
\end{abstract}

\maketitle

Since the discovery of the intrinsic Josephson effect in highly
anisotropic layered Bi$_2$Sr$_2$CaCu$_2$O$_{8+x}$ (Bi-2212)
high-$T_c$ single crystals \cite{Kleiner}, tunneling
characteristics along the $c$ axis have been extensively
investigated using the mesa structure prepared on the crystal
surface to probe the interlayer coupling characteristics as well
as the superconducting properties of the Cu-O layers themselves
\cite{mesa}. Superconducting properties of Bi-2212 crystals have
also been examined using tunneling in artificial surface junctions
or the scanning tunneling spectroscopy \cite{Renner}. In this
case, however, tunneling properties are susceptible to any surface
degradation. The advantage of tunneling measurements using
intrinsic junctions, in comparison, is that one can eliminate the
surface-dependent effect.

The poor thermal conductivity of the Bi-2212 intrinsic junctions,
however, is known to cause serious local self-heating in tunneling
measurements, when a high-density bias current is used
\cite{Krasnov,heat}. The resulting temperature variation often
causes a serious spurious effect in the tunneling signal. Much
effort has been made to avoid self-heating by reducing the lateral
size and the thickness of the mesa, the contact resistance, and by
employing the pulsed bias method \cite{mesa,pulse}. Taking these
precautions, one can reduce the back-bending effect (negative
dynamics resistance) in the current-voltage ($I-V$) curve, which
is mainly caused by self-heating in a high-bias range. In-situ
temperature measurements in the above configurations, however,
show that, even in the absence of the back bending, the
self-heating still persists and can significantly distort
tunneling measurements. \cite{ratio,Yurgens}

In this letter we present a scheme of making tunneling
measurements in a stack of intrinsic junctions at a constant
temperature in any finite bias currents. Our samples consisted of
two separate stacks of intrinsic junctions which were closely
coupled laterally by a common Au electrode [inset of Fig. 1(b)].
We monitored the temperature of the stack of interest [``sample
stack (SmS)"] including self-heating, which was then compensated
using the proportional-integral-derivative (PID) temperature
control method with another stack of intrinsic junctions
[``thermometer stack (ThS)"] as a thermometer and using a
substrate-holder heater.

Bi-2212 single crystals were grown by the solid-state-reaction
method \cite{Kim}. In this study, two samples [SH1 and SH2] were
fabricated, where the double-side-cleaving of Bi-2212 crystals,
micropatterning, and ion-beam etching were employed \cite{Bae}.
Resistive transition of the two samples exhibited that SH1 was
slightly underdoped ($T_c$=85.8 K, $x$=0.21) and SH2 was almost
optimally doped ($T_c$=90.3 K, $x$=0.23) \cite{Suzuki}. The inset
of Fig. 1(b) shows the sample geometry and the measurement
configuration. The left and right stacks are the SmS and the ThS,
respectively. The lateral dimensions of the SmS in SH1 and SH2
were 15.3$\times$1.6 and 16.2$\times$1.6 $\mu$m$^2$, respectively.
The ThS was placed laterally 1.4 and 1.6 $\mu$m apart from the SmS
for SH1 and SH2, respectively, and was thermally coupled by a
100-nm-thick common Au electrode [inset of Fig. 1(b)]. Contrary to
the case of a mesa structure, without the poorly
thermal-conductive Bi-2212 basal part in this geometry, the heat
generated in the SmS diffuses effectively through the highly
thermal-conductive ($\kappa$=100 W/m$\cdot$K) bottom Au electrode,
thus putting both the SmS and ThS in the thermal equilibrium
presumably at an almost equal temperature. The tunneling
characteristics of the SmS and the ThS were measured in two- and
three-terminal configurations, respectively, with the Au common
ground electrode.

\begin{figure}[t]
\begin{center}
\leavevmode
\includegraphics[width=0.8\linewidth]{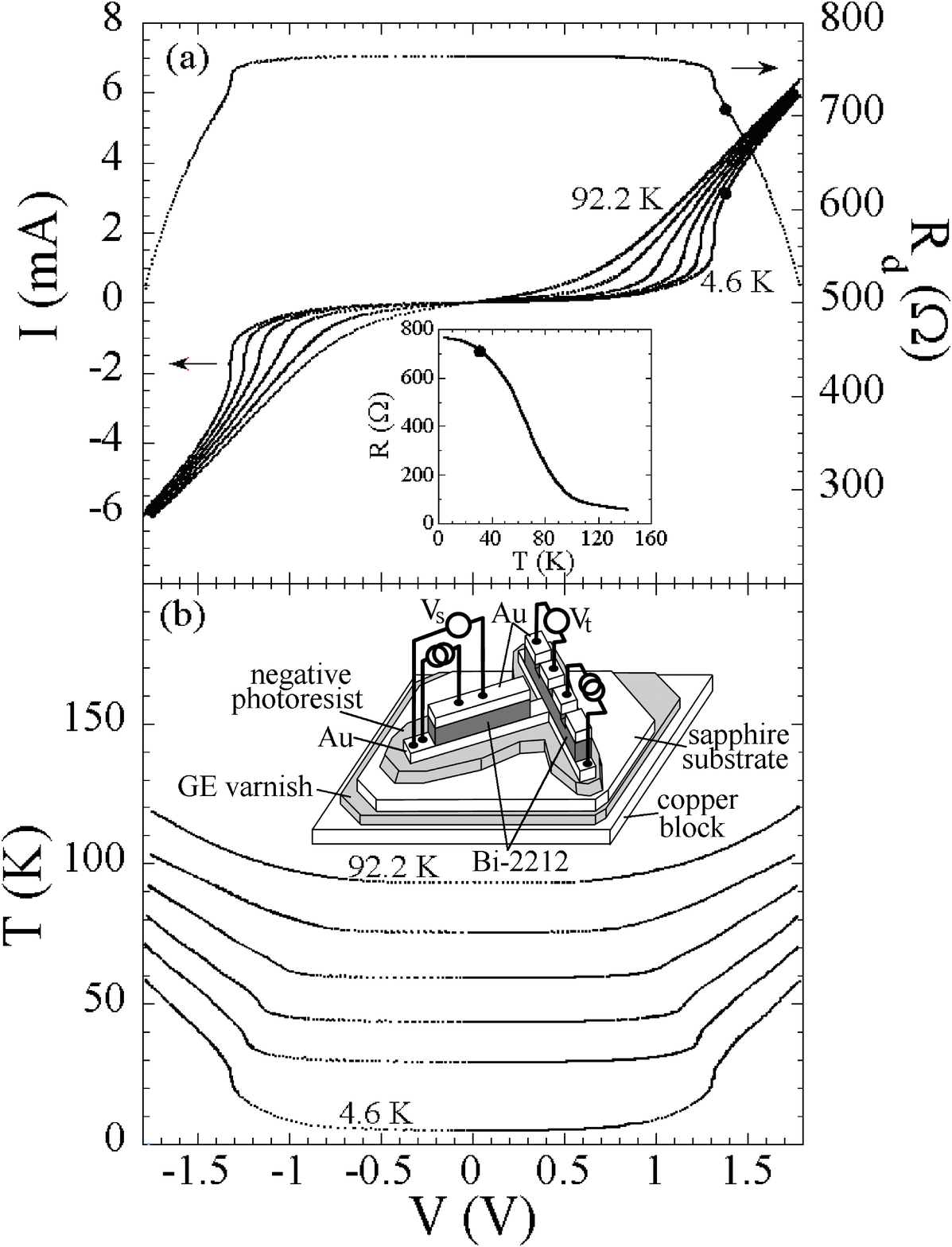}
\caption{(a) $I-V$ characteristics of the sample stack (SmS) in
SH1 for the substrate-holder temperature at 92.2, 75.1, 58.9,
43.5, 29.2, and 4.6 K, sequentially. The upper set of data is the
resistance of the thermometer stack (ThS) at the substrate-holder
temperature of 4.6 K in the bias of 0.1 mA, while the bias current
of the SmS is swept. Inset: The $R$ vs $T$ curve of the ThS in the
bias of 0.1 mA. (b) The temperature of the SmS corresponding to
the $I-V$ curves in (a) as a function of bias voltage. Inset: the
sample geometry and the measurement configuration.}
\end{center}
\end{figure}

In our SmS without the basal part the temperature is almost
uniformly distributed along the $c$ axis. For instance, the SmS of
SH1 (containing $N$=25 intrinsic junctions) was 37.5 nm thick.
According to the thermal-conduction relation, $\Delta T$=$j_c
Vt/\kappa _c$, with the top junction in the normal state in a high
bias current density $j>j_c$, the temperature difference between
top and bottom of the stack is predicted to be only $\sim$ 0.8 K.
In this estimate, the $c$-axis critical current density of
$j_c$=0.6 kAcm$^{-2}$, the voltage over each junction of $V$=30
mV, and the $c$-axis thermal conductivity of $\kappa_c$=0.009
W/m$\cdot$K at $T$=4.2 K, were used \cite{Krasnov}. If all the
junctions in the SmS are driven to the normal state the whole SmS
acts as a heating element and the temperature will become more
uniform.

\begin{figure}[b]
\begin{center}
\leavevmode
\includegraphics[width=0.75\linewidth]{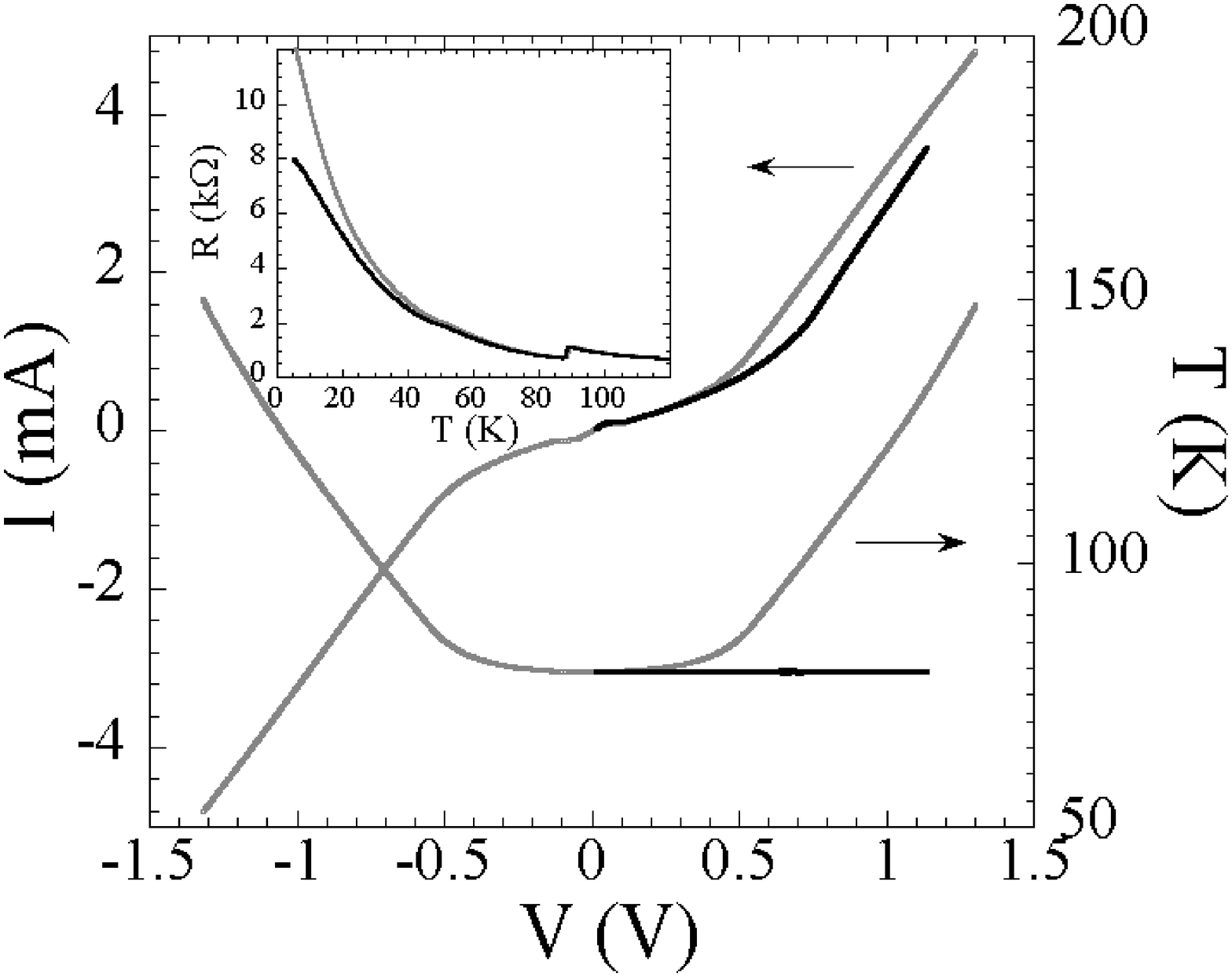}
\caption{(a) $I-V$ characteristics and temperatures of the sample
stack as a function of the bias voltage without (gray curves) and
with (black curves) the PID temperature control scheme. Inset: The
$R$ vs $T$ curves of the sample stack near zero bias without (gray
curve) and with (black curve; only the positive-bias data were
taken) the bias current of 0.28 mA in the thermometer stack.}
\end{center}
\end{figure}

The temperature increase in the ThS due to self-heating of the SmS
reduces the gap energy and the quasiparticle tunneling resistance
of the ThS. The inset of Fig. 1(a) shows the $R$ vs $T$ curve of
the ThS of SH1 in the bias current of 0.1 mA, which is slightly
above its tunneling critical current at 4.6 K. The temperature of
the ThS was monitored by measuring the variation of the
quasiparticle tunneling resistance in the bias of 0.1 mA, while
the bias current of the SmS was swept continuously. To illustrate
the temperature monitoring scheme, one first notices in Fig. 1(a)
that the bias current of 2.8 mA, for instance, corresponds to the
voltage value of 1.3 V in the SmS as denoted by a dot on the $I-V$
curve at 4.6 K. As illustrated with the right vertical scale,
under these circumstances, the quasipariticle resistance of the
ThS was reduced from 760 $\Omega$, corresponding to $T$=4.6 K, to
720 $\Omega$ because of the heat flow from the SmS. According to
the $R$ vs $T$ curve in the inset of Fig. 1(a), the reduced
resistance indicates that, due to self-heating, the temperature of
both stacks increased from 4.6 K to 30 K.

Fig. 1(b) displays the variation of the actual temperature of the
SmS, determined in the way described above along with the $I-V$
curves shown in Fig. 1(a), for different biases and the
substrate-holder temperatures. In principle, the bias current of
0.1 mA in the ThS may also generate self-heating. The $I-V$
characteristics of the SmS with and without a bias current in the
ThS, however, showed no noticeable difference between the two
cases (data not shown). In our samples the averaged heating ratio
($i.e.$, the temperature increase per dissipated power) was $\sim$
5 K/mW at 4.2 K, which was at least an order lower than the
previously reported value of $\sim$ 50 K/mW in the mesa structure
\cite{ratio, Yurgens}.

Although the elimination of the basal Bi-2212 part and the strong
thermal coupling through the Au ground electrode reduced the
heating ratio the self-heating was not completely eliminated. The
main cause of the self-heating was the poor thermal conductivity
of the gluing materials used. Negative photoresist was used to fix
the Au-sandwiched Bi-2212 stack to the sapphire substrate, which
in turn was attached to the copper substrate holder by GE varnish.
The thermal conductivity of GE varnish, 0.08 W/m$\cdot$K (at
$T$=10 K), is much smaller than that of sapphire of 20
W/m$\cdot$K. The photoresist is expected to have even poorer
thermal conductivity than GE varnish.

To illustrate the usefulness of our PID temperature control scheme
we performed the tunneling spectroscopy using SH2 at fixed
temperatures. The PID system consisted of a manganin heater coiled
around the substrate holder, where the feedback current was
determined by the relation \cite{Chan},
\begin{equation}
I(t)=P\Delta T(t)+\frac{P}{\tau}\int^{t}_{-\infty} \Delta
T(t')dt'+DP\frac{d\Delta T(t)}{dt}.
\end{equation}
Here, $\Delta T(t)$ is the difference between the set temperature
and the actual temperature of the stacks. And $P$ is the
proportional gain, $\tau$ the integration time constant, and $D$
the time rate constant. To monitor the temperature of the SmS of
SH2, using the $R$ vs $T$ curves of the ThS, the ThS was biased by
0.28 mA, which was slightly above the tunneling critical current
of SH2. As illustrated in the inset of Fig. 2, the higher bias
current in SH2 than in SH1, caused serious self-heating. The gray
and black solid lines in the inset are the $R$ vs $T$ curves of
the SmS ($I_{sample}$= 0.5 $\mu$A) in zero current and 0.28 mA in
the ThS, respectively. Using these two $R$ vs $T$ curves we
eliminated any additional error in reading temperatures due to the
bias current of 0.28 mA in SH2.

\begin{figure}[t]
\begin{center}
\leavevmode
\includegraphics[width=0.7\linewidth]{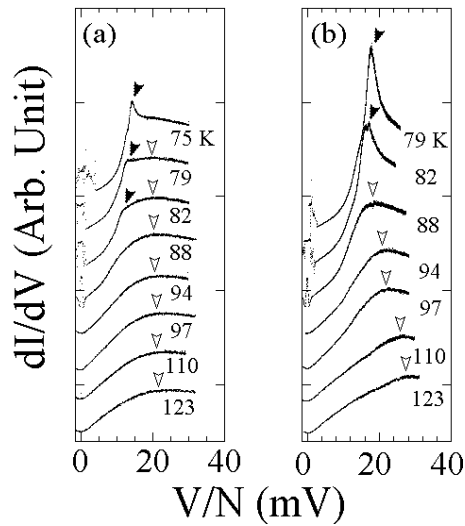}
\caption{Dynamic tunneling conductance (a) without and (b) with
the PID temperature control for varying substrate-holder
temperature near $T_c$. Arrows are explained in the text.}
\end{center}
\end{figure}

Fig. 2 shows the $I-V$ curves of SH2 without (the upper gray
curve) and with (the upper black curve) adopting the PID scheme.
The sweep speed of the bias in the SmS with the PID scheme was
$\sim$50 $\mu$Hz, much slower than 1 mHz without the PID scheme.
The lower gray and black curves in Fig. 2 illustrate the
temperature variation of the SmS as a function of the voltage in
the cases without and with the PID scheme, respectively. Without
the PID scheme, for the substrate-holder and bath temperatures at
79 K and 4.2 K, respectively, the temperature corresponding to the
bias voltage of 1 V was $\sim$120 K. Using the PID scheme with the
parameters $P$=20, $\tau$=10, and $D$=1.2, however, the
temperature of the SmS was kept fixed at 79$\pm$0.4 K for all the
biases used. For higher biases further decrease of the
substrate-holder temperature is required to keep the SmS
temperature constant, which limits the varying ranges of the bias
and the substrate-holder temperature. The window of the varying
range of the two parameters shrinks for lower set temperatures of
the SmS. For this reason, in this study, the tunneling
measurements could be done for temperatures only above $\sim$40 K.

Fig. 3 shows the temperature dependence of the dynamical tunneling
conductance (obtained by the lock-in technique) of SH2 near $T_c$.
Here, the bias voltage values were normalized by the number of
junctions, $N$=44. The open arrows in the figure indicate the
maximum positions of the humps, which can be used to estimate the
size of the pseudogap. On the other hand, the filled arrows
indicate the superconducting gap edges. As reported previously
\cite{mesa} and in Fig. 3(a), without employing the PID scheme,
the maximum positions of the humps are almost insensitive to
temperature variation. As in Fig. 3(b), however, when the
temperature of the stack was kept fixed by employing the the PID
scheme for a set of conductance data, the maximum positions of the
humps increase with increasing temperature above $T_c$. In
addition, without the PID scheme the superconducting gap and
pseudogap have separate values near $T_c$ (93.5 K) of SH2, but
they tend to merge into one with the PID scheme. The implication
of these results should be reexamined.

In conclusion, we devised a PID scheme to keep the temperature of
stacks of intrinsic junctions constant in a bias current by
compensating the temperature increase of the stacks due to
self-heating with lowering the substrate-holder temperature. We
demonstrated, with an example of tunneling-spectroscopic
measurements, that eliminating the self-heating is essential in
obtaining intrinsic tunneling properties for any high biases in
highly anisotropic layered cuprates such as Bi-2212.

The authors wish to acknowledge valuable discussion with V. M.
Krasnov in Chalmers University of Technology as well as Y.-B. Kim
and M.-S. Shin in POSTECH. This work was supported by the National
Research Laboratory project administrated by KISTEP and also by
the AOARD of the US Air Force by the Contract No.
FA5209-04-P-0253.

\end{document}